\documentclass[journal,twoside,web]{ieeecolor}
\usepackage{tmi}
\usepackage{cite}
\usepackage{amsmath,amssymb,amsfonts}
\usepackage{algorithmic}
\usepackage{graphicx}
\usepackage{textcomp}
\usepackage{dsfont}
\usepackage{color, soul}

\def\BibTeX{{\rm B\kern-.05em{\sc i\kern-.025em b}\kern-.08em
    T\kern-.1667em\lower.7ex\hbox{E}\kern-.125emX}}
\usepackage{graphicx,amssymb,mathrsfs,amsmath,amsfonts,dsfont}
\usepackage{hyperref}
\usepackage{lineno}
\usepackage{stfloats}
\usepackage{algorithm}
\usepackage{algorithmic}
\usepackage{amsbsy}
\usepackage[mathscr]{euscript}
\usepackage{bm}
\usepackage{multirow}
\usepackage[utf8]{inputenc}

\newcommand{\x}{\bm{x}}

\newcommand{\y}{\bm{y}}
\newcommand{\z}{\bm{z}}

\newcommand{\n}{\bm{n}}


\newtheorem{pro}{Proposition}[section]

\newtheorem{rmk}{Remark}[section]
\newtheorem{assump}{Assumption}

\begin{document}
\title{Self-Score: Self-Supervised Learning on Score-Based Models for MRI Reconstruction}
\author{Zhuo-Xu~Cui, Chentao~Cao, Shaonan~Liu, Qingyong~Zhu, Jing~Cheng, Haifeng~Wang, Yanjie~Zhu, Dong~Liang, \IEEEmembership{Senior Member, IEEE}
\thanks{This work was supported in part by the National Key R$\&$D Program of China (2020YFA0712202, 2017YFC0108802 and 2017YFC0112903); China Postdoctoral Science Foundation under Grant 2020M682990; National Natural Science Foundation of China (61771463, 81830056, U1805261, 81971611, 61871373, 81729003, 81901736); Natural Science Foundation of Guangdong Province (2018A0303130132); Shenzhen Key Laboratory of Ultrasound Imaging and Therapy (ZDSYS20180206180631473); Shenzhen Peacock Plan Team Program (KQTD20180413181834876); Innovation and Technology Commission of the government of Hong Kong SAR (MRP/001/18X); Strategic Priority Research Program of Chinese Academy of Sciences (XDB25000000).}
\thanks{Corresponding author:dong.liang@siat.ac.cn}
\thanks{Z.-X. Cui and C. Cao contributed equally to this work.}
\thanks{Z.-X. Cui, Q. Zhu and D. Liang are with Research Center for Medical AI, Shenzhen Institutes of Advanced Technology, Chinese Academy of Sciences, Shenzhen, China.}
\thanks{C. Cao, J. Cheng, H. Wang, Y. Zhu and D. Liang are with Paul C. Lauterbur Research Center for Biomedical Imaging, Shenzhen Institutes of Advanced Technology, Chinese Academy of Sciences, Shenzhen, China.}
\thanks{S. Liu is with Department of Computer Science, Inner Mongolia University, Hohhot, China.}
}

\maketitle

\begin{abstract}
Recently, score-based diffusion models have shown satisfactory performance in MRI reconstruction. Most of these methods require a large amount of fully sampled MRI data as a training set, which, sometimes, is difficult to acquire in practice. This paper proposes a fully-sampled-data-free score-based diffusion model for MRI reconstruction, which learns the fully sampled MR image prior in a self-supervised manner on undersampled data. Specifically, we first infer the fully sampled MR image distribution from the undersampled data by Bayesian deep learning, then perturb the data distribution and approximate their probability density gradient by training a score function. Leveraging the learned score function as a prior, we can reconstruct the MR image by performing conditioned Langevin Markov chain Monte Carlo (MCMC) sampling. Experiments on the public dataset show that the proposed method outperforms existing self-supervised MRI reconstruction methods and achieves comparable performances with the conventional (fully sampled data trained) score-based diffusion methods.
\end{abstract}


\section{Introduction}
\label{sec:introduction}

\IEEEPARstart{A}{ccurate} reconstruction of MR images from undersampled measurements is at the heart of accelerated MRI. Mathematically, MRI reconstruction can be reduced to solving an inverse problem. In the early years, traditional methods, including parallel imaging \cite{pruessmann1999sense,Griswold2002Generalized,Lustig2010spirit} and compressed sensing \cite{Candes2006Robust,Donoho2006Compressed,Lustig2007Sparse}, were usually used to solve MRI inverse problems by introducing hand-crafted priors to realize regularization. However, as the MRI acceleration is further improved, the performance of the methods based on hand-crafted prior degrades \cite{8962949}.
Inspired by the tremendous success of deep learning (DL), DL methods based on data-driven prior characterization have received considerable attention for MRI reconstruction \cite{7493320,yang2016deep,zhu2018Image,8756028,Huang2021Deep,9481093,9481108}.

Recently, score-based diffusion DL models, including denoising score-matching models \cite{NEURIPS2019_3001ef25,NEURIPS2020_92c3b916}, denoising diffusion probabilistic models (DDPMs) \cite{NEURIPS2020_4c5bcfec} and unified stochastic differential equation (SDE) models \cite{song2021scorebased}, etc., have received much attention. Specifically, the score-based diffusion model forward process gradually perturbs the real data to random noise, and the probability density function on the diffusion trajectory is characterized by training a score function. Leveraging the learned score function as a prior, we can perform the Langevin Markov chain Monte Carlo (MCMC) sampling to generate the desired data from random noise. Focusing on MRI, if the score-based diffusion model forward process gradually perturbs the fully sampled MR images, its conditioned Langevin MCMC sampling enables MRI reconstruction. Related studies \cite{NEURIPS2021_7d6044e9,chung2022score} have shown that the scored-based diffusion model outperforms existing conventional DL imaging methods regarding MRI reconstruction accuracy and generalization ability.

The score function must be trained on a fully sampled MRI dataset for existing scored-based diffusion models. However, accessing numerous fully sampled data might be challenging. Thanks to the development of self-supervised learning in computer vision \cite{lehtinen2018noise2noise,Moran_2020_CVPR,Quan_2020_CVPR}, some models have been extended to MRI reconstruction. In particular, no fully sampled dataset is required, and these models learn MR reconstruction mappings only from undersampled data \cite{9098514,millard2022self,9721153}. However, these self-supervised learning methods are often inferior to supervised learning methods in terms of reconstruction accuracy. Additionally, self-supervised learning methods usually do not generalize well, e.g., models trained under a certain sampling trajectory perform poorly when applied to other sampling trajectories. As mentioned above, the score-based diffusion model has advantages in terms of reconstruction accuracy and generalization. Therefore, it is desired to propose a self-supervised learning score-based diffusion model with high reconstruction accuracy and high generalization capability for MRI reconstruction.

\subsection{Contributions and Observations }
Motivated by the abovementioned problems, this paper proposes a new self-supervised DL method for MRI reconstruction. Specifically, the main contributions and observations of this work are summarized as follows.
\begin{enumerate}
	\item We propose a self-supervised learning score-based diffusion model for the scenario without a fully sampled MRI training set and derive the corresponding conditioned Langevin MCMC sampling for MRI reconstruction.
	\item In the proposed model, we show that the score of fully sampled MR image probability density can be estimated accurately under certain conditions depending only on the undersampled data. It provides theoretical guarantees for the proposed self-supervised learning MRI reconstruction.
\item In terms of reconstruction accuracy and generalization ability (including sampling trajectories and data shifts), experimental results show that our proposed method outperforms traditional parallel imaging, self-supervised DL, and conventional supervised DL methods and achieves comparable performance with conventional (fully sampled data trained) score-based diffusion methods.
\end{enumerate}

The remainder of the paper is organized as follows. Section \ref{sect_rw} reviews some related work. Section \ref{sect3} discusses the proposed self-supervised score-based diffusion model. The implementation details are presented in Section \ref{sect4}. Experiments performed on several data sets are presented in Section \ref{sect5}. The discussions are presented in Section \ref{sect6}. The last section \ref{sect7} gives some concluding remarks. All the proofs are presented in the Appendix.

\section{Related Work}\label{sect_rw}
\subsection{Score-Based Diffusion Models}
Suppose a certain data set $\{\x_i\in \mathbb{C}^d\}_{i=1}^{N}$ contains i.i.d. sampled data from an unknown distribution $p(\x)$ and the score function $\bm{s}_{\bm{\phi}}:\mathbb{C}^d\rightarrow\mathbb{C}^d$ with parameter $\bm{\phi}$ is an approximation of the gradient of its log probability density, i.e., $\nabla \log p(\x)$.
Then, Langevin MCMC sampling is performed according to the score function to obtain samples that obey the distribution $p(\x)$, i.e.,
\begin{equation}\label{mcmc}
\x_{i+1}=\x_{i}+\frac{\eta_i}{2}\bm{s}_{\bm{\phi}}(\x_{i})+\sqrt{\eta_i}\z_t
\end{equation}
where $\eta_i>0$ is the stepsize, and $\z_t$ is standard normal. $\nabla \log p(\x)$ is difficult to calculate directly, the denoising score matching method \cite{vincent2011connection} first perturbs the original data $\x$ to $\widetilde{\x}$, and the score function on $\widetilde{\x}$ can be obtained by solving the following optimization problem:
\begin{equation*}
\min_{\bm{\phi}}\mathbb{E}_{q(\widetilde{\x},\x)}\left[\frac{1}{2}\left\|\bm{s}_{\bm{\phi}}(\widetilde{\x})-\frac{\partial \log q_{\varepsilon}(\widetilde{\x}|\x)}{\partial \widetilde{\x}}\right\|^2\right]\\
\end{equation*}
where $q_{\varepsilon}(\widetilde{\x}|\x):=\mathcal{N}(\widetilde{\x};\x,\varepsilon^2\bm{I})$. Given a sequence of positive noise scales $\varepsilon_{\min}=\varepsilon_1<\cdots<\varepsilon_L=\varepsilon_{\max}$, the data noise perturbation process can be considered as a discretization of a continuous diffusion process. In particular, several works model this diffusion process through Markov chains \cite{NEURIPS2020_4c5bcfec} and It\^{o} SDEs \cite{song2021scorebased}. Leveraging the learned score function, we can perform the reverse process (Langevin MCMC sampling) of above diffusion model to generate the data $\widetilde{\x}\sim q_{\varepsilon_{\min}}(\widetilde{\x})$ from random noise ${\n}\sim q_{\varepsilon_{\max}}({\n})$, where $q_{\varepsilon_{i}}(\cdot)=\int q_{\varepsilon_{i}}(\cdot|\x)p(\x)\text{d}\x$. Typically, if $\varepsilon_{\min}$ is sufficiently small, it can be assumed that $q_{\varepsilon_{\min}}(\cdot)\approx p(\cdot)$.

Focusing on MRI, data $\{\x_i\in \mathbb{C}^d\}_{i=1}^{N}$ represents the fully sampled MR images. In many practical applications, a large training set of fully sampled MRI images is difficult to acquire, which limits the application of score-based diffusion models.
\subsection{Self-Supervised Learning Methods}
Self-supervised learning has been studied in MRI for a longer period. For example, the classical $k$-space interpolation method, which first fully samples a calibrated region in the central part of the $k$-space, learns a linear kernel and uses it in a translation-invariant manner to interpolate in the missing $k$-space data \cite{Griswold2002Generalized,Lustig2010spirit}. Inspired by deep learning, linear kernels are generalized to convolutional neural networks to improve the accuracy of missing data interpolation \cite{akccakaya2019scan,kim2019loraki}.

On the one hand, inspired by self-supervised learning denoising methods in computer vision, self-supervised MRI reconstruction methods that do not require fully sampled calibration data have been proposed \cite{9098514,millard2022self,cui2022k}. More specifically, an undersampled $k$-space data pair is constructed by drawing on the self-supervised denoising method to construct a noisy data pair, and then the interpolation relationship between the missing data and the sampled data is learned from it to achieve missing data interpolation (MRI reconstruction).

However, in terms of imaging quality, many experiments show that self-supervised learning methods tend to perform inferior to supervised learning methods. Moreover, as in $k$-space interpolation methods, the self-supervised learning of the interpolation kernel relies on the sampling trajectory, thus leading to a lack of generalization ability. Therefore, a novel self-supervised learning method with high reconstruction quality and generalization ability has been desired for MRI reconstruction.

\section{Methodology}\label{sect3}
In this section, we first estimate MR image data distribution by self-supervised Bayesian learning and then introduce the probability density gradient estimation by the score matching method to perform Langevin MCMC sampling (MRI reconstruction).
\subsection{MRI Forward Model}
The forward model of MRI can be expressed as
\begin{equation}\label{eq:1}\y=\bm{A}\x+\n\end{equation}
where $\bm{A}$ is the encoding matrix, $\y$ is the measurement, $\x$ is the image to be reconstructed, and $\n$ is the measurement Gaussian noise with scare $\gamma$ i.e., $\n\sim \mathcal{N}(\n;0,\gamma^2\bm{I})$.
In particular, $\bm{A=PFS}$ for the case of multichannel acquisition, where $\bm{S}$ denotes the coil sensitivities and $\bm{F}$ represents the Fourier transform, and $\bm{P}$ denotes the undersampling pattern.
Since $A$ is an ill-conditioned operator, it is difficult to reconstruct the image $\x$ from $\y$ accurately and stably. Therefore, it is necessary to introduce a prior $p(\x)$ of $\x$ to realize regularization. This paper's main task is to accurately estimate the distribution $p(\x)$ of the fully sampled image $\x$ from the undersampled data $\y$.
\begin{figure*}[thbp]
\begin{center}
\includegraphics[width=0.95\textwidth,height=0.25\textwidth]{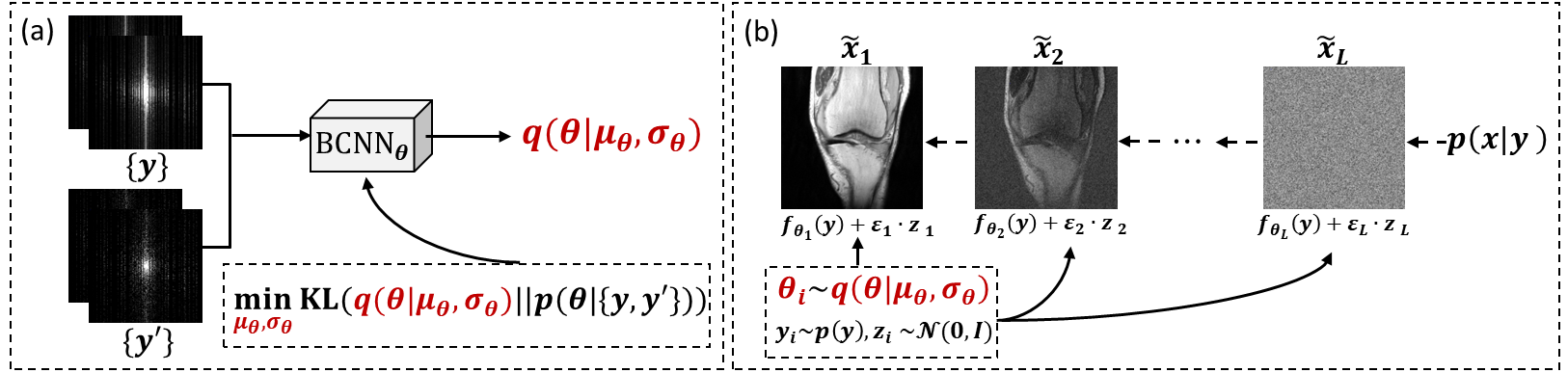}
\end{center}
\caption{Illustration of the self-supervised score-based diffusion model. (a) first, the undersampled $k$-space data pairs $\{\y,\y'\}$ are constructed, the BCNN is trained by minimizing the KL divergence, and finally the model parameter distribution $q(\bm{\theta}|\bm{\mu_{\theta}},\bm{\sigma_{\theta}})$ is output. Specifically, the the architecture of BCNN follows the POCS-SPIRiT model driven neural network \cite{cui2021equilibrated} and its network parameters contain standard deviation $\bm{\sigma}$ and mean $\bm{\mu}$. When tested and trained, a normal distribution noise $\bm{\epsilon}\sim \prod_s\mathcal{N}(\epsilon_s|0,1)$ is first generated and then the convolutional kernel is determined by sampling, i.e., $\bm{\theta}=\bm{\mu}+\bm{\sigma}\bm{\epsilon}$. (b) Forward process: learn the score function approximating the probability density gradient of $\x$ by perturbing the $\bm{f_{\theta}(\y)}$ with Gaussian noise at different scales. Reverse process: perform MCMC sampling conditional on the measurement $\y$ to reconstruct MR image using the learned score function as a prior.}
\label{f1}
\end{figure*}
\subsection{Self-Supervised Bayesian Learning}
Before presenting our method, we make the following key assumption:
\begin{assump}\label{ass:1}
Let $\x$ denote the MR image, $\y$ denote the measurement, and $\y'$ denotes a further subsample of $\y$. Suppose there exists a mapping $\bm{f}_{\bm{\theta}}$ with parameter $\bm{\theta}$ such that
\begin{equation*}
\bm{f}_{\bm{\theta}}(\y)=\x+\n_1 ~\text{and}~ \bm{f}_{\bm{\theta}}(\y')=\widehat{\y}+\n_2
\end{equation*}
where $\widehat{\y}$ denotes the image obtained by inverse Fourier transform and channel merging of $\y$, i.e., $\widehat{\y}=\bm{S^*F^{-1}y}$, $\n_1$ and $\n_2$ denote the Gaussian noise with scales $\gamma_1$ and $\gamma_2$.
\end{assump}
\begin{rmk}
This assumption can be considered a generalization of the "linear interpolability" and "translation invariance" of the classical $k$-space interpolation method. In addition, it was proved theoretically that this mapping exists by expectation \cite{millard2022self}.
\end{rmk}

According to Assumption \ref{ass:1}, we can obtain the conditional distribution $p(\x|\y,\bm{\theta})$, and if we can further obtain the distribution of the parameters $\bm{\theta}$, we can infer the distribution of image $\x$ using the following Bayesian formula
\begin{equation}\label{p_x}p(\x)=\int p(\x|\y,\bm{\theta})p(\y)p(\bm{\theta})\text{d}\y\text{d}\bm{\theta}.\end{equation}
Therefore, our next objective is to estimate the distribution of $p(\bm{\theta})$. Defining undersampled $k$-space data pairs $D:=\{(\y_i,{\y'}_i)\}_{i=1}^N$, we estimate $p(\bm{\theta}|D)$ by Bayesian inference, i.e., by using $q(\bm{\theta}|\bm{\mu_{\theta}},\bm{\sigma_{\theta}})$ to estimate $p(\bm{\theta}|D)$, where
$$q(\bm{\theta}|\bm{\mu_{\theta}},\bm{\sigma_{\theta}})\sim \prod_{s}\mathcal{N}(\theta_s|\mu_{\theta,s},\sigma_{\theta,s})$$
$\bm{\theta}=\{\theta_s\}_s$, $\bm{\mu_{\theta}}=\{\mu_{\theta,s}\}_s$, and $\bm{\sigma_{\theta}}=\{\sigma_{\theta,s}\}_s$.
In this paper, we use a Bayesian convolutional neural network (BCNN) to represent $\bm{f_{\theta}}$ and output the distribution $q(\bm{\theta}|\bm{\mu_{\theta}},\bm{\sigma_{\theta}})$. The framework of self-supervised Bayesian DL is shown in Figure \ref{f1}. In particular, the architecture of mapping $\bm{f_{\theta}}$ or BCNN can be directly adopted from the POCS-SPIRiT model driven neural network \cite{cui2021equilibrated}. Introduced randomness, the network parameters contain standard deviation $\bm{\sigma_{\theta}}$ and mean $\bm{\mu_{\theta}}$. For parameter $\bm{\theta}$, we first generate normal distribution noise $\bm{\epsilon}\sim\prod_{s}\mathcal{N}(\epsilon_s|0,1)$, and then determine the
convolutional kernel by sampling, i.e., $\bm{\theta} = \bm{\mu_{\theta}} +\bm{\sigma_{\theta}}\bm{\epsilon}$.

The tight approximation is achieved by minimizing the following KL-divergence:
\begin{equation}\label{kl}
\min_{\bm{\mu_\theta,\sigma_\theta}}\text{KL}(q(\bm{\theta}|\bm{\mu_{\theta}},\bm{\sigma_{\theta}})||p(\bm{\theta}|D)).
\end{equation}
The final calculation result is as follows:
\begin{pro}\label{prop:1}
Suppose that $p(\n_2)\sim \prod_t \mathcal{N}(n_t|0,\gamma_2)$ and prior $p(\bm{\theta})\sim \prod_s\mathcal{N}(\theta_s|0,\overline{\sigma}_\theta)$. The minimum of the KL-divergence (\ref{kl}) can be calculated by
\begin{equation}\begin{aligned}\label{prior4}
&\min_{\bm{\mu_\theta,\sigma_\theta}}\frac{1}{2\gamma_2^2}\mathbb{E}_{q(\bm{\theta}|\bm{\mu_\theta,\sigma_\theta})}\sum_{i=1}^N\|\bm{f_\theta}({\y'}_i)-\y_i\|^2\\
&+\frac{1}{2\overline{\sigma}_\theta^2}(\|\bm{\mu_\theta}\|^2+\|\bm{\sigma_\theta}\|^2)-\sum_s\log\left(\frac{\sigma_{\theta,i}}{\overline{\sigma}_\theta}\right) + \text{const}.
\end{aligned}\end{equation}
\end{pro}
The calculation process of Proposition \ref{prop:1} is shown in Appendix \ref{app:1}.
\subsection{Score-Based Diffusion Model}
Bringing the above estimated $q(\bm{\theta}|\bm{\mu_\theta,\sigma_\theta})$ into the Bayesian equation (\ref{p_x}), we can obtain the distribution $p(\x)$. However, the integral in (\ref{p_x}) is difficult to calculate exactly in general.
If $\nabla \log p(\x)$ is accessible, samples obeying the distribution $p(\x)$ can be collected according to Langevin MCMC sampling (\ref{mcmc}).

Following the score matching method, we can estimate $\nabla \log p(\x)$ in a self-supervised manner by the following optimization problem.
\begin{pro}\label{prop:2}
Suppose Assumption \ref{ass:1} holds. The minimizer of $\mathbb{E}_{p({\x})}[1/2\|\bm{s}_{\bm{\phi}}(\x)-\nabla \log p(\x)\|^2]$ can be obtained by equivalently minimizing the following objective:
$$\min_{\bm{\phi}}\mathbb{E}_{p({\x},\y,\bm{\theta})}\left[\frac{1}{2}\left\|\bm{s}_{\bm{\phi}}({\x})-\frac{\partial \log p({\x}|\bm{\theta},\y)}{\partial {\x}}\right\|^2\right].$$
\end{pro}
The calculation process of Proposition \ref{prop:2} is shown in Appendix \ref{app:2}.

Based on Proposition \ref{prop:2}, the score function at different perturbation levels can be learned, and then the desired samples are obtained by performing MCMC sampling according to them.
More specifically, we perturb the $\bm{f_{\theta}(\y)}$ by Gaussian noise with scales $\{\varepsilon_i\}_{i=1}^T$ that satisfies $\varepsilon_1<\varepsilon_2<\cdots<\varepsilon_T$. Let $p_{\varepsilon_i}(\widetilde{\x},|\y,\bm{\theta}))=\mathcal{N}(\widetilde{\x}|\bm{f_{\theta}(\y)},\varepsilon_i^2\bm{I})$ and perturbed data distribution is $p_{\varepsilon_i}(\widetilde{\x})=\int p_{\varepsilon_i}(\widetilde{\x}|\y,\bm{\theta})p(\y)p(\bm{\theta})\text{d}\y\text{d}\bm{\theta}$. If $\varepsilon_1$ is chosen such that $\varepsilon_1=\gamma_1$, then $p_{\varepsilon_1}(\x)=p(\x)$ holds. Based on Proposition \ref{prop:2}, we can estimate $\nabla \log p_{\varepsilon_i}(\widetilde{\x})$ at all the scales by training a joint score function $\bm{s}_{\bm{\phi}}(\widetilde{\x},\varepsilon_i)$ with the following loss:
\begin{equation}\label{score_match}\frac{1}{2L}\sum_{i=1}^{L}\mathbb{E}_{p(\y)q(\bm{\theta})}\mathbb{E}_{p_{\varepsilon_i}(\widetilde{\x},|\y,\bm{\theta})}\left[\left\|\varepsilon_i\bm{s}_{\bm{\phi}}(\widetilde{\x},\varepsilon_i)+\frac{\widetilde{\x}-\bm{f_{\theta}(\y)}}{\varepsilon_i}\right\|^2\right]\end{equation}
where $q(\bm{\theta}):=q(\bm{\theta}|\bm{\mu_\theta,\sigma_\theta})$ is the minimizer of objective (\ref{prior4}).

After training an score function, we can perform the condition Langevin MCMC sampling to solve problem (\ref{eq:1}), i.e.,
\begin{equation*}\begin{aligned}
\x_{i+1}&=\x_{i}+\frac{\eta_i}{2}\nabla \log p(\x_{i}|\y)+\sqrt{\eta_i}\z_i\\
&=\x_{i}+\frac{\eta_i}{2}(\nabla \log p(\x_{i})+\nabla \log p(\y|\x_{i}))+\sqrt{\eta_i}\z_i\\
&=\x_{i}+\frac{\eta_i}{2}\left(\bm{s}_{\bm{\phi}}(\x_{i},\varepsilon_{i})+\frac{\bm{A^*}(\bm{A}\x_{i}-\y)}{\gamma^2+\varepsilon_i^2}\right)+\sqrt{\eta_i}\z_i
\end{aligned}\end{equation*}
In particular, the choice of parameter $\eta_i$ follows that of literature \cite{NEURIPS2020_92c3b916}, and the conditional Langevin MCMC sampling is detailed in Algorithm \ref{alg:1}.
\begin{algorithm}[htb]
	\caption{Conditional Langevin MCMC Sampling.}
	\label{alg:1}
	\begin{algorithmic}[1]
		\STATE {\bfseries Input:} $\{\varepsilon_i\}_{i=1}^L$, $\epsilon$ and $T$;\\
		\STATE {\bfseries Initialize:} $\widetilde{\x}_0$;\\
		\FOR{$i=L,L-1,\ldots,1$}
        \STATE $\eta_i=\epsilon\cdot\varepsilon_i^2/\varepsilon^2_L$
        \FOR{$t=0,2,\ldots,T-1$}
		\STATE Draw $\z\sim \mathcal{N}(\bm{0},\bm{I})$:
\begin{small}\begin{equation*}
\widetilde{\x}_{t+1}=\widetilde{\x}_{t}+\frac{\eta_i}{2}\left(\bm{s}_{\bm{\phi}}(\widetilde{\x}_{t},\varepsilon_{i})+\frac{\bm{A^*}(\bm{A}\widetilde{\x}_{t}-\y)}{\gamma^2+\varepsilon_i^2}\right)+\sqrt{\eta_i}\z_t
\end{equation*}\end{small}
		\ENDFOR
        \STATE $\widetilde{\x}_0=\widetilde{\x}_T$
        \ENDFOR
		\STATE {\bfseries Output:} $\widetilde{\x}_T$.
	\end{algorithmic}
\end{algorithm}

The advantages of our proposed method over existing self-supervised learning methods (learning the mapping $\bm{f_{\theta}}$ directly from undersampled data) are twofold.
\begin{enumerate}
\item Intuitively, the score matching model (\ref{score_match}) works by separating out the noise from the current image $\widetilde{\x}$, so even if $\bm{f_{\theta}(\y)}$ is not a perfectly clean image, the impact on the noise separation mechanism ($\bm{s}_{\bm{\phi}}(\cdot)$) learning is relatively small.
\item Methodologically, we utilize the prior $p(\x)=\int p(\x|\y,\bm{\theta})p(\y)q(\bm{\theta})\text{d}\y\text{d}\bm{\theta}$ on a set of model $\bm{\theta}$ over a set of data $\y$ rather than the prior of a single model over a single set of data i.e., $\bm{f_{\theta_i}(\y_i)}$, $\bm{\theta}_i \sim q(\bm{\theta})$ and $\y_i\sim p(\y)$. It has been shown that such ensemble models can often outperform single models \cite{zhou2012ensemble}.
\end{enumerate}

\section{Implementation}\label{sect4}
The evaluation was performed on fastMRI public MRI data, and the data acquired by our Siemens scans with various $k$-space trajectories. The details of the MR data are as follows:
\subsection{Data Acquisition}
\subsubsection{FastMRI data}
 The knee raw data \footnote{\url{https://fastmri.org/}} was acquired from a 3T Siemens scanner (Siemens Magnetom Skyra, Prisma and Biograph mMR). Data acquisition used a 15 channel knee coil array and conventional Cartesian 2D TSE protocol employed clinically at NYU School of Medicine. The following sequence parameters
were used: Echo train length 4, matrix size $320 \times 320$, in-plane resolution $0.5mm\times0.5mm$, slice thickness $3mm$, no gap between slices. Timing varied between systems, with repetition time (TR) ranging between 2200 and 3000 milliseconds, and echo time (TE) between 27 and 34 milliseconds. From them, we randomly select T1-weighted data of 34 individuals (1002 slices in total) as the training set and data of 3 individuals (95 slices in total) as the test set.

To verify the generalizability, we will test the knee data trained model's performance on brain MRI reconstruction.
Therefore, a randomly selected set of brain (10 slices in total) data from the fastMRI dataset is also used as the training set.
\subsubsection{SIAT data}
For the training data, overall 1000 fully sampled multi contrast data from 10 subjects with a 3T scanner (Siemens Trio, Siemens AG, Erlangen, Germany) were collected and informed consent was obtained from the
imaging object in compliance with the IRB policy. The fully sampled data was acquired by a 12-channel head coil with matrix size of $256 \times 256$ and combined to single-channel complex-valued data. For the testing set, we draw 50 slices from the data of 7 human brain datasets acquired from three different commercial 3T scanners (SIEMENS AG, Erlangen, Germany; GE Healthcare, Waukesha, WI; United Imaging Healthcare, Shanghai, China).

\subsection{Network Architecture and Training}
\begin{figure}[!t]
\centering
\includegraphics[width=0.48\textwidth,height=0.14\textwidth]{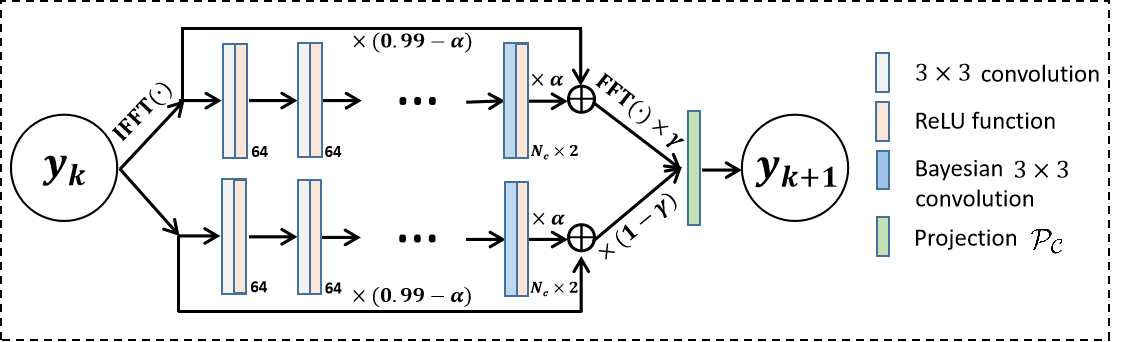}
\caption{Schematic diagram of the network architecture of the $\bm{f_{\theta}}$. The upper and lower convolutional network modules exploit redundancies in the image domain and self-consistency in the $k$-space. $\mathcal{P}_{\mathcal{C}}$ denotes the projection onto $\y'$, i.e., $\mathcal{P}_{\mathcal{C}}(\x)=\bm{(I-P')}\x+\y'$, where $\bm{P'}$ is the undersampling pattern of $\y'$. In particular, the above module contains five (Bayesian) convolutional layers and recurses it ten times to obtain the architecture of $\bm{f_{\theta}}$. In particular, $\bm{f_\theta}$, as well as the distribution of $\bm{\theta}$, constructs BCNN.}
\label{f_net}
\end{figure}

The schematic diagram of the BCNN (including $\bm{f_{\theta}}$) architecture is illustrated in Figures \ref{f_net}. Since Assumption \ref{ass:1} stems from the generalization of "linear predictability" and "flat invariance" of the $k$-space data, we followed the SPIRiT-POCS model neural network \cite{cui2021equilibrated} for representing function $\bm{f_{\theta}}$. In particular, to introduce randomness, we replaced the last layer of the module with a Bayesian neural network. On the other hand, we used the NCSNv2 network \cite{NEURIPS2020_92c3b916} to learn the score function, whose specific parameters are set: $\varepsilon_L=50$, $\varepsilon_1=0.0066$, number of classes is 266, ema is true, ema rate is 0.999.

\begin{figure}[!t]
\centering
\includegraphics[width=0.3\textwidth,height=0.18\textwidth]{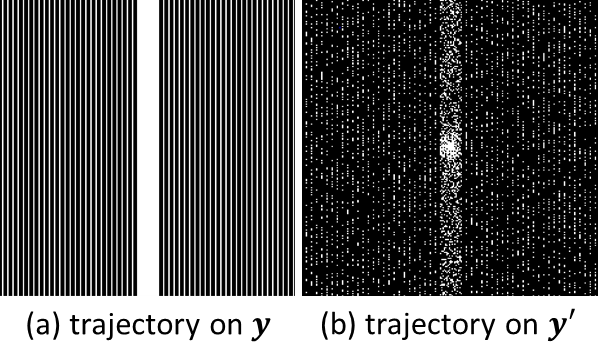}
\caption{An example of an undersampling trajectory on $\y$ and an undersampling trajectory on $\y'$. }
\label{f_mask}
\end{figure}

The ADAM \cite{kingma2014adam} optimizer with $\beta_1=0.9, \beta_2=0.999$ is chosen for optimizing loss functions (\ref{score_match}) and (\ref{prior4}). The size of the mini batch is 1, the number of epochs is 200 and the learning rate is $10^{-4 }$.
As shown in Figure \ref{f_mask}, in constructing undersampled data pairs, we followed \cite{9098514} to obtain $\y'$ by further Gaussian undersampling of $\y$.
The models were implemented on an Ubuntu 20.04 operating system equipped with an NVIDIA A6000 Tensor Core (GPU, 48 GB memory) in the open PyTorch 1.10 framework \cite{paszke2019pytorch} with CUDA 11.3 and CUDNN support.

\subsection{Performance Evaluation}
In this study, the quantitative evaluations were all calculated on the image domain. The reconstructed and reference images were derived using an inverse Fourier transform followed by an elementwise square-root of sum-of-the squares (SSoS) operation, i.e. $\z[n]=(\sum_{i=1}^{N_c}|\x_i[n]|^2)^{\frac{1}{2}}$, where $\z[n]$ denotes the $n$-th element of image $\z$, and $\x_i[n]$ denotes the $n$-th element of the $i$th coil image $\x_i$. For quantitative evaluation, the peak signal-to-noise ratio (PSNR), normalized mean square error (NMSE) value, and structural similarity (SSIM) index \cite{1284395} were adopted.

\section{Experimentation Results}\label{sect5}
In this section, a series of extensive comparative experiments were studied. In particular, we compared the traditional PI method SENSE \cite{pruessmann1999sense}. Since the BCNN (termed "self-supervised") proposed in Section \ref{sect3} can be considered an improved version of the self-supervised MRI reconstruction method SSDU \cite{9098514}, we made it a comparison method and an ablation method.
To further verify the superiority of the proposed method, we also compared it with the supervised MRI reconstruction method ISTA-Net (termed "supervised") \cite{Zhang_2018_CVPR}. Specifically, our code is available at \url{https://github.com/ZhuoxuCui/Self-Score}.

\subsection{In-Distribution Performance}
\begin{figure*}[!t]
\centering
\includegraphics[width=0.96\textwidth,height=0.65\textwidth]{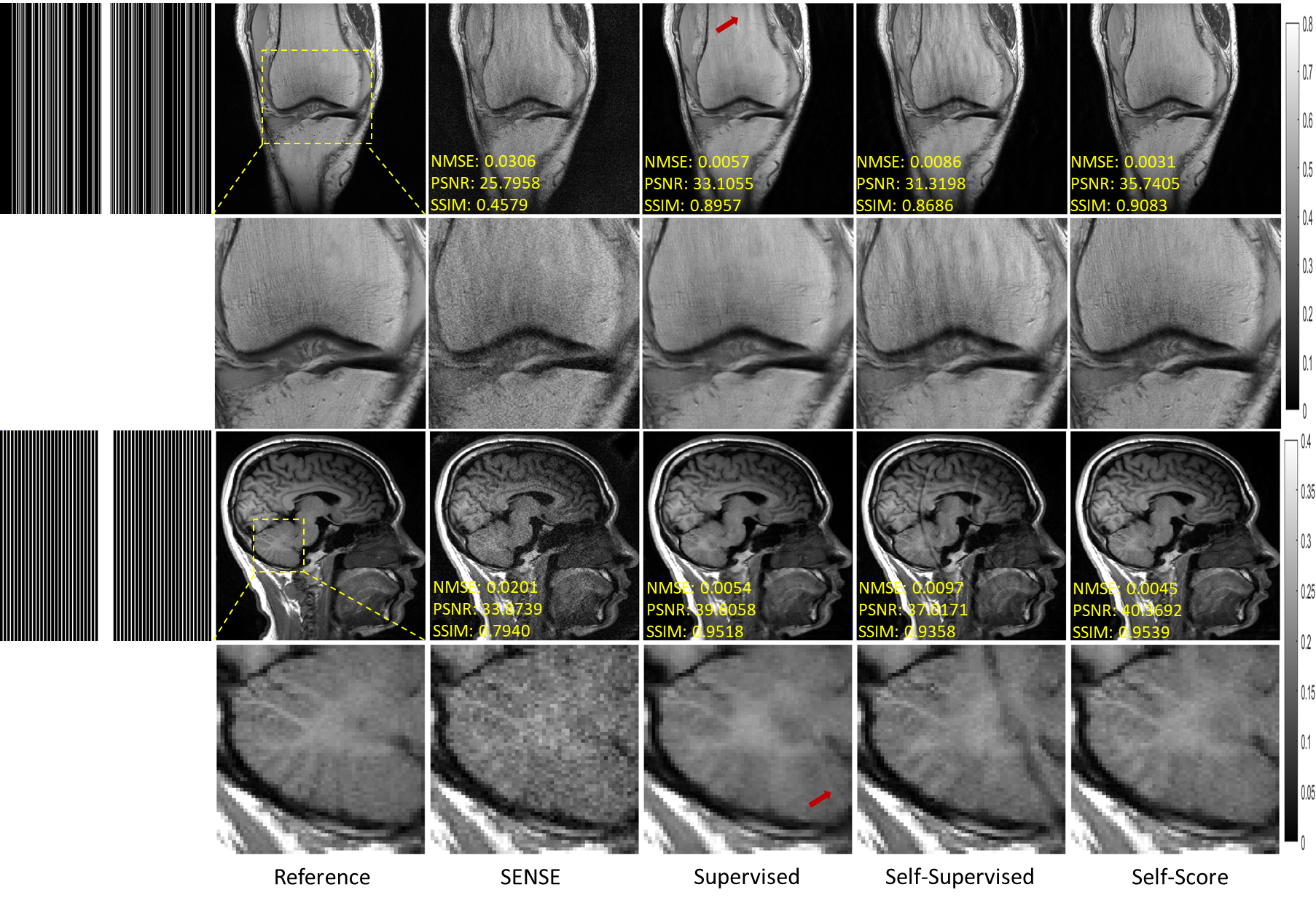}
\caption{Reconstruction of the fastMRI knee (first two rows) and SIAT brain (last two rows) data at random sampling of $R=5$ and uniform undersampling of $R=4$, respectively. The values in the corner are NMSE/PSNR/SSIM values. Second and fourth rows illustrate the enlarged views. The grayscale of the reconstructed images is at the right of the figure.}
\label{f2}
\end{figure*}

In this section, we tested the performance of various methods when sampling patterns and data types (anatomies) are consistent in training and testing. Figure \ref{f2} shows the reconstruction results for 5-fold random undersampling of fastMRI knee data and 4-fold uniform sampling of SIAT brain data. For the traditional PI method SENSE, the aliasing pattern remains in the reconstructed image and noise is amplified to the point that image details are obliterated. An obvious aliasing pattern remains in the reconstructed images for the self-supervised learning method. Although the noise is well suppressed for the supervised learning method, a tiny aliasing pattern remains in the reconstructed image where the red arrow points. The reconstructed image is also blurrier, and some texture information is lost from the visual perception. Our proposed method performs well in aliasing pattern suppression and image texture detail recovery. In particular, it is worth mentioning that the proposed fully-sampled-data-free method outperforms the conventional supervised DL method that requires a fully sampled training dataset, which is of practical significance.

The competitive quantitative results of the above methods are shown in Table \ref{tab:1a}. Our method consistently outperforms traditional SENSE, conventional self-supervised and supervised methods characterized by quantitative evaluations. The above experiments confirm our method's competitiveness under consistent training and testing environments.

\begin{table}
	\begin{center}
		\caption{Quantitative comparison for various methods on the fastMRI and SIAT dataset.}\label{tab:1a}
		\setlength{\tabcolsep}{1.0mm}{
			\begin{tabular}{l|l|ccc}
				\hline
				\multicolumn{ 2}{c}{ Datasets} & \multicolumn{ 3}{|c}{Quantitative Evaluation}  \\
				\multicolumn{ 2}{c|}{ \& Methods   } &NMSE &PSNR(dB)&SSIM  \\
				\hline
				\multirow{4}{*}{fastMRI Knee}
				& SENSE  &0.0412$\pm$0.0457&27.01$\pm$2.96&0.52$\pm$0.14 \\
				\cline{2-5}
				& Supervised  &0.0072$\pm$0.0053&34.00$\pm$2.06&0.90$\pm$0.03 \\
				\cline{2-5}
				& Self-Supervised  &0.0089$\pm$0.0036&32.77$\pm$1.45&0.88$\pm$0.02 \\
				\cline{2-5}
				& Self-Score  &\textcolor{red}{0.0037$\pm$0.0012}&\textcolor{red}{36.46$\pm$1.70}&\textcolor{red}{0.91$\pm$0.02}\\
				\hline
				\multirow{4}{*}{SIAT Brain}
				& SENSE  &0.0312$\pm$0.0136&31.40$\pm$1.57&0.71$\pm$0.05 \\
				\cline{2-5}
				& Supervised  &0.0061$\pm$0.0021&38.25$\pm$1.55&0.94$\pm$0.01 \\
				\cline{2-5}
				& Self-Supervised  &0.0088$\pm$0.0022&36.54$\pm$1.24&0.94$\pm$0.01 \\
				\cline{2-5}
				& Self-Score  &\textcolor{red}{0.0051$\pm$0.0016}&\textcolor{red}{39.04$\pm$1.08}&0.94$\pm$0.01\\
				\hline
		\end{tabular}}
	\end{center}
\end{table}

\subsection{Out-of-Distribution Performance}
Good generalization ability is necessary for realistic applications of DL methods. However, as mentioned earlier, some existing self-supervised learning MR reconstruction methods are generalized from $k$-space interpolation, which will not work when sampling patterns and $k$-space data structures are inconsistent in training and testing. The proposed method eventually learns the distribution of fully sampled MR images, free from sampling patterns and data structure limitations. Therefore, the proposed method tends to have better generalization ability. To verify this claim, we designed two experiments as follows.
\subsubsection{Pattern Shift}
\begin{figure*}[!t]
\centering
\includegraphics[width=0.96\textwidth,height=0.6\textwidth]{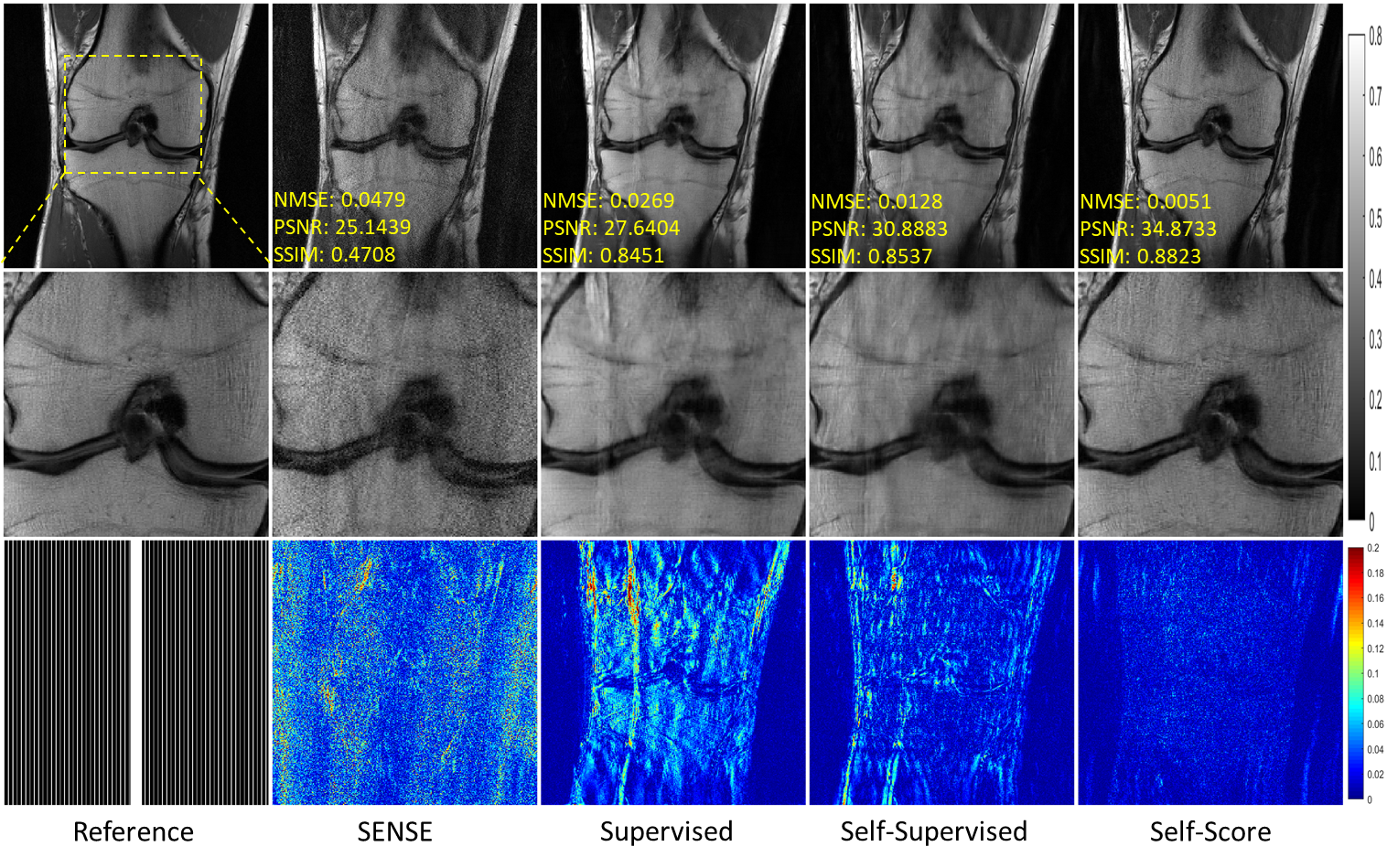}
\caption{Reconstruction results under uniform undersampling at $R=6$. The corresponding models were trained on a randomly sampled pattern at $R=5$. The values in the corner are NMSE/PSNR/SSIM values of each slice. Second and third rows illustrate the enlarged and error views, respectively. The grayscale of the reconstructed images and the color bar of the error images are at the right of the figure.}
\label{f3}
\end{figure*}

In this section, we verified the performance of various methods when the sampling patterns were inconsistent during training and testing. Figure \ref{f3} shows the results of various methods trained on a 5-fold randomly sampling pattern and reconstructed on 6-fold uniformly sampled data. Although SENSE is not affected by the pattern shift, it performs poorly because the prior is not sufficiently introduced. The aliasing pattern remains in the reconstruction results of both supervised and self-supervised learning methods. In particular, comparing Figure \ref{f2} with Figure \ref{f3}, we can see that both supervised and self-supervised methods degrade significantly due to the pattern shift. On the other hand, it is easy to see that our proposed method achieves satisfactory performance in both aliasing pattern suppression and detail recovery. The quantitative metrics are shown in Table \ref{tab:1b}, confirming our method's superiority.

\begin{table}
	\begin{center}
		\caption{Quantitative comparison for various methods on the fastMRI knee dataset.}\label{tab:1b}
		\setlength{\tabcolsep}{1.0mm}{
			\begin{tabular}{l|l|ccc}
				\hline
				\multicolumn{ 2}{c}{ Datasets} & \multicolumn{ 3}{|c}{Quantitative Evaluation}  \\
				\multicolumn{ 2}{c|}{ \& Methods   } &NMSE &PSNR(dB)&SSIM  \\
				\hline
				\multirow{4}{*}{fastMRI Knee}
				& SENSE  &0.0529$\pm$0.0535&25.84$\pm$2.97&0.48$\pm$0.14 \\
				\cline{2-5}
				& Supervised  &0.0152$\pm$0.0091&30.67$\pm$2.06&0.88$\pm$0.02 \\
				\cline{2-5}
				& Self-Supervised  &0.0171$\pm$0.0094&30.19$\pm$2.09&0.87$\pm$0.02 \\
				\cline{2-5}
				& Self-Score  &\textcolor{red}{0.0049$\pm$0.0023}&\textcolor{red}{35.49$\pm$1.54}&\textcolor{red}{0.90$\pm$0.03}\\
				\hline
		\end{tabular}}
	\end{center}
\end{table}

The above experiments validate the superior generalization ability of the proposed method in terms of sampling pattern shifts.
\subsubsection{Data Shift}

\begin{figure*}[!t]
\centering
\includegraphics[width=0.8\textwidth,height=0.4\textwidth]{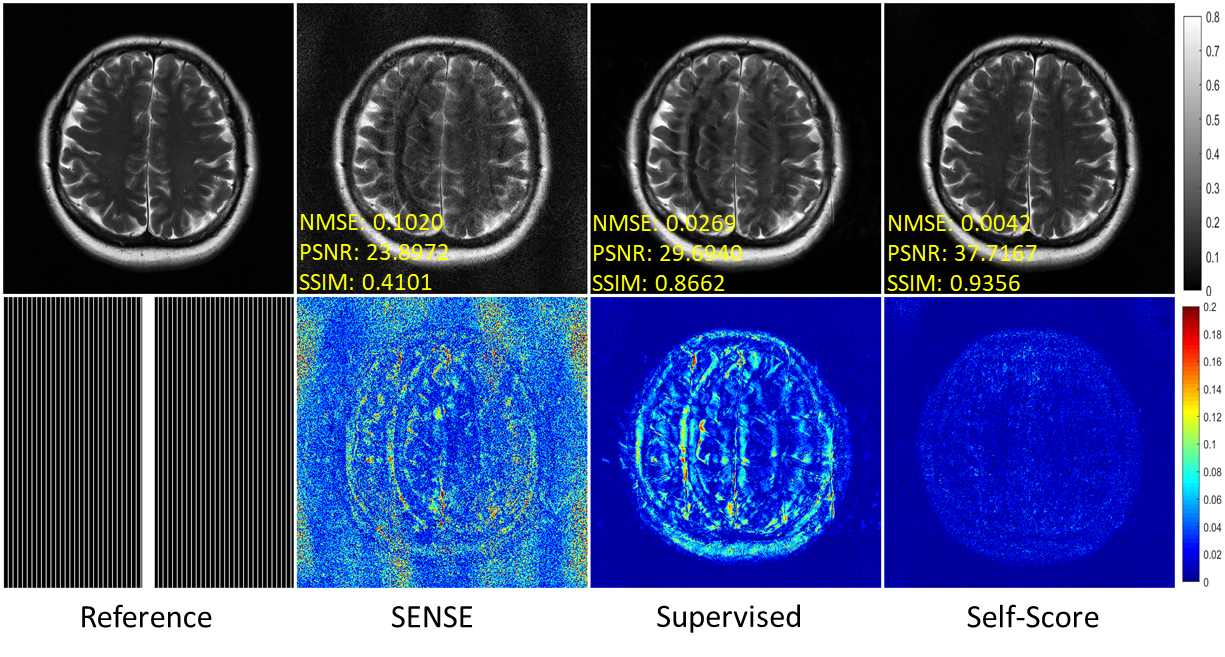}
\caption{Reconstruction results of fastMRI brain data under uniform undersampling at $R=6$. The corresponding models were trained on fastMRI knee data. The values in the corner are NMSE/PSNR/SSIM values of each slice. Second and third rows illustrate the enlarged and error views, respectively. The grayscale of the reconstructed images and the color bar of the error images are at the right of the figure.}
\label{f4}
\end{figure*}

In this section, we verified the performance of various methods when the data type (anatomies) were inconsistent during training and testing. Figure \ref{f4} shows the results of various methods trained on fastMRI knee data and reconstructed on fastMRI brain. The self-supervised method based on $k$-space interpolation in the paper is no longer applicable due to the different number of channels in the two data sets. Similar to the pattern shift experiment results, SENSE performs poorly, although it is not affected by data shift. The supervised learning method reconstructs the image aliasing pattern residuals, verifying that it also generalizes poorly in data shift. The proposed method in this paper can accurately reconstruct images, thus verifying its superior generalization in data shift.

\begin{table}
	\begin{center}
		\caption{Quantitative comparison for various methods on the fastMRI brain dataset.}\label{tab:1c}
		\setlength{\tabcolsep}{1.0mm}{
			\begin{tabular}{l|l|ccc}
				\hline
				\multicolumn{ 2}{c}{ Datasets} & \multicolumn{ 3}{|c}{Quantitative Evaluation}  \\
				\multicolumn{ 2}{c|}{ \& Methods   } &NMSE &PSNR(dB)&SSIM  \\
				\hline
				\multirow{3}{*}{fastMRI Brain}
				& SENSE  &0.1703$\pm$0.1559&22.05$\pm$1.56&0.31$\pm$0.09 \\
				\cline{2-5}
				& Supervised  &0.0273$\pm$0.0124&29.37$\pm$1.22&0.85$\pm$0.03 \\
				\cline{2-5}
				& Self-Score  &\textcolor{red}{0.0068$\pm$0.0042}&\textcolor{red}{35.53$\pm$1.31}&\textcolor{red}{0.90$\pm$0.03}\\
				\hline
		\end{tabular}}
	\end{center}
\end{table}

The above experiments verified that the proposed method can accurately reconstruct images and has good generalization ability without the fully sampled training data.
\section{Discussion}\label{sect6}
In this study, we proposed a self-supervised score-based diffusion model for MRI reconstruction. In the above comparative experiments, we verified the superiority of the proposed method compared to the traditional parallel imaging, self-supervised DL and conventional supervised DL methods. However, some areas still need further discussion or improvement for our proposed model.
\subsection{Comparison Experiment with Conventional Score-Based Method}
Theoretically, the accuracy of the proposed method regarding the estimation of data distribution depends mainly on Assumption \ref{ass:1}. Although it is a natural generalization of "linear predictability" and "translational invariance" in $k$-space interpolation methods, there may still be some cases where it is not fully accurate. Therefore, we designed comparison experiments with the conventional score-based diffusion method trained on fully sampled data to verify the accuracy of the proposed self-supervised learning method on the data distribution estimation.

\begin{figure}[!t]
\centering
\includegraphics[width=0.48\textwidth,height=0.62\textwidth]{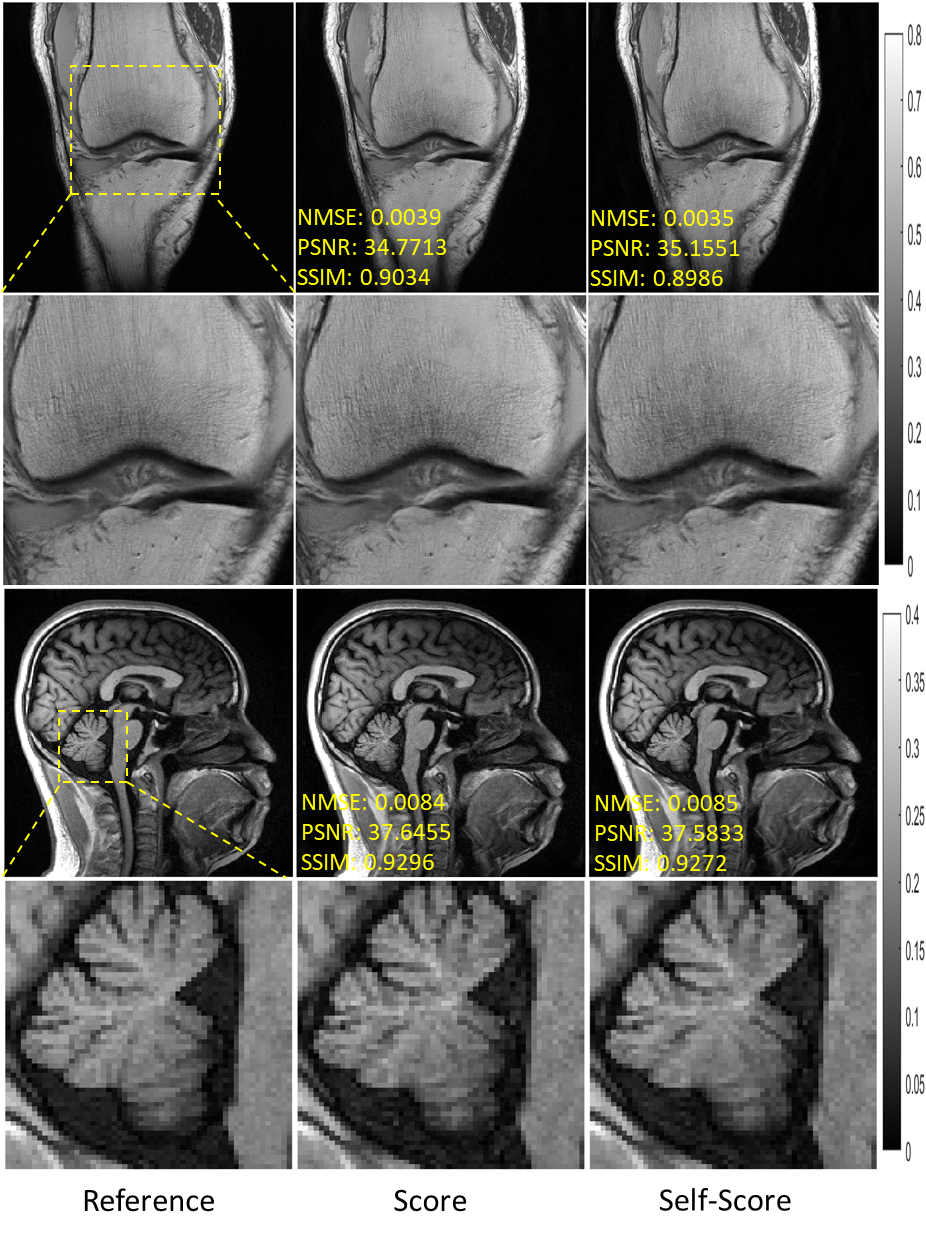}
\caption{Reconstruction of the fastMRI knee (first two rows) and SIAT brain (last two rows) data at uniform undersampling of $R=6$ and $R=4$, respectively. The values in the corner are NMSE/PSNR/SSIM values. Second and fourth rows illustrate the enlarged views. The grayscale of the reconstructed images is at the right of the figure.}
\label{f5}
\end{figure}

Figure \ref{f5} shows the reconstruction results of the conventional score-based method (termed score) and our proposed self-supervised score-based method (termed self-score) for fastMRI knee and SIAT brain data under 6-fold and 4-fold uniform undersampling, respectively. In terms of visual perception, the two methods perform almost identically.
The quantitative metrics are shown in Table \ref{tab:2}. It can be found that the proposed self-score method performs almost identically to the score method on the fastMRI knee dataset and even slightly better than the score on the SIAT brain dataset.
This experiment validates the accuracy of the proposed self-supervised learning method for data distribution estimation.

\begin{table}
	\begin{center}
		\caption{Quantitative comparison for various methods on the fastMRI and SIAT dataset.}\label{tab:2}
		\setlength{\tabcolsep}{1.6mm}{
			\begin{tabular}{l|l|ccc}
				\hline
				\multicolumn{ 2}{c}{ Datasets} & \multicolumn{ 3}{|c}{Quantitative Evaluation}  \\
				\multicolumn{ 2}{c|}{ \& Methods   } &NMSE &PSNR(dB)&SSIM  \\
				\hline
				\multirow{2}{*}{fastMRI Knee}
				& Score  &\textcolor{red}{0.0048$\pm$0.0027}&\textcolor{red}{35.64$\pm$1.78}&0.90$\pm$0.04 \\
				\cline{2-5}
				& Self-Score  &0.0049$\pm$0.0023&35.49$\pm$1.54&\textcolor{red}{0.90$\pm$0.03}\\
				\hline
				\multirow{2}{*}{SIAT Brain}
				& Score &0.0057$\pm$0.0019&38.57$\pm$1.17&0.92$\pm$0.03 \\
				\cline{2-5}
				& Self-Score  &\textcolor{red}{0.0051$\pm$0.0016}&\textcolor{red}{39.04$\pm$1.08}&\textcolor{red}{0.94$\pm$0.02}\\
				\hline
				\multirow{2}{*}{fastMRI Brain}
				& Score &\textcolor{red}{0.0066$\pm$0.0054}&\textcolor{red}{36.10$\pm$1.64}&0.90$\pm$0.04 \\
				\cline{2-5}
				& Self-Score  &0.0068$\pm$0.0042&35.53$\pm$1.31&\textcolor{red}{0.90$\pm$0.03}\\
				\hline
		\end{tabular}}
	\end{center}
\end{table}

\begin{figure}[!t]
\centering
\includegraphics[width=0.48\textwidth,height=0.34\textwidth]{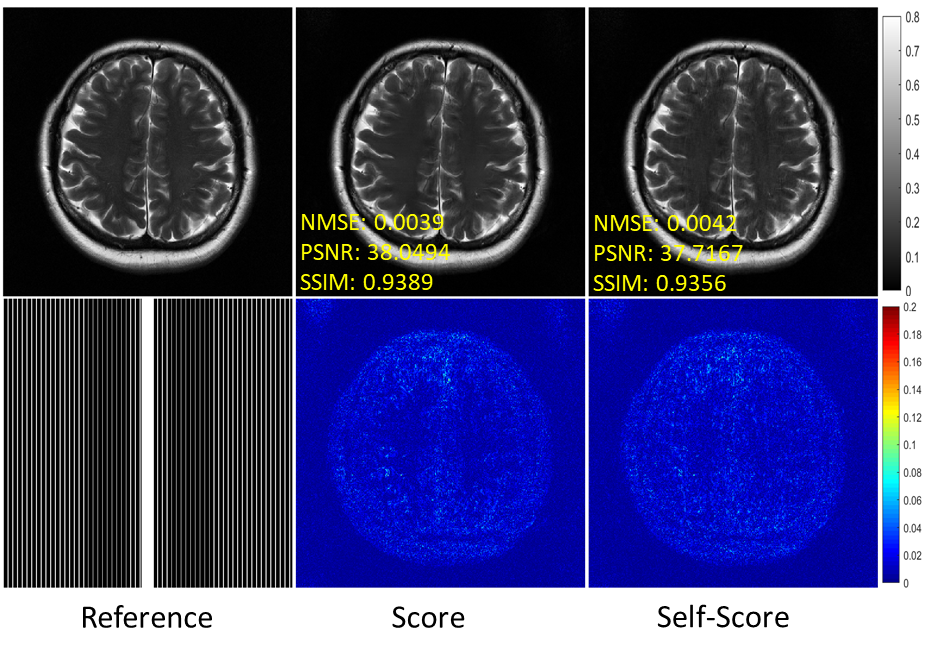}
\caption{Reconstruction of the fastMRI brain at uniform undersampling of $R=6$. The values in the corner are NMSE/PSNR/SSIM values. Second row illustrates the error view. The grayscale of the reconstructed images and the color bar of the error images are at the right of the figure.}
\label{f6}
\end{figure}

We also compared the generalization capabilities of the two methods described above, i.e., trained on the fastMRI knee dataset but applied to reconstruct 6-fold uniformly undersampled fastMRI brain MR images. From the results presented in Figure \ref{f6}, we can find that the proposed self-supervised score-based method trained on undersampled data performs very close to the method trained on fully sampled data in terms of the visual perception and error of the reconstructed images. The above experimental result further validates the accuracy of the proposed self-supervised learning method for estimating the distribution of the fully sampled MRI data.

\subsection{Further Discussion and Improvements}
As mentioned above, although Assumption \ref{ass:1} is a common assumption in $k$-space interpolation methods, there is no rigorous theory to guarantee its correctness. Therefore, it is a future work for us to design a self-supervised score-based method for MRI reconstruction without relying on Assumption \ref{ass:1}.

For the parameter $\bm{\theta}$ distribution estimation, we use a simple Gaussian distribution $q(\bm{\theta}|\bm{\mu_{\theta}},\bm{\sigma_{\theta}})$ to approximate it. On the other hand, it is also possible to collect the sample $\bm{\theta}_i$ by MCMC sampling and then approximate the distribution of $\bm{\theta}$ by $\sum_{i}\delta_{\bm{\theta}=\bm{\theta}_i}$. Given the successful application of $q(\bm{\theta}|\bm{\mu_{\theta}},\bm{\sigma_{\theta}})$ approximation in computer vision \cite{pang2020self,li2022supervised}, such an approach is adopted in this paper. It is worth noting that the MCMC method and the more accurate distribution approximation methods will be reserved as our options.

Finally, for conventional DL MRI reconstruction methods, the learned network can map the undersampled $k$-space data directly to the reconstructed image, which takes little time. Our method (including other score-based diffusion methods) needs to perform an iteration (MCMC sampling) to reconstruct the image, which takes a relatively long time. Reducing the reconstruction (sampling) time by introducing acceleration means, such as momentum, is a direction worth investigating in the future.

\section{Conclusion}\label{sect7}
This paper proposed a self-supervised score-based diffusion model for MRI reconstruction. Unlike the conventional score-based method, the proposed method learns the full-sampling MRI image prior to the undersampled data in a self-supervised manner. Based on the learned prior, this paper also demonstrated the implementation of MRI reconstruction with MCMC sampling conditional on undersampled $k$-space data. Experiments on the public dataset showed that the proposed method outperforms existing self-supervised and conventional supervised MRI reconstruction methods and achieves comparable performances with the conventional (fully sampled data trained) score-based diffusion methods. Our method could be a powerful framework for MRI reconstruction, and further development of this method may enable even larger gains in the future.
\appendix
\subsection{Proof of Proposition \ref{prop:1}}\label{app:1}
By Bayesian formula, the KL-divergence can be calculated as
\begin{equation*}\begin{aligned}
&\text{KL}(q(\bm{\theta}|\bm{\mu_\theta},\bm{\sigma_\theta})\|p(\bm{\theta}|D))\\
=&\text{KL}(q(\bm{\theta}|\bm{\mu_\theta},\bm{\sigma_\theta})\|p(\bm{\theta}))-\mathbb{E}_{q(\bm{\theta}|\bm{\mu_\theta},\bm{\sigma_\theta})}\log p(D|\bm{\theta})+\text{const}.
\end{aligned}\end{equation*}
For the first part, we have
\begin{equation*}\begin{aligned}
&\text{KL}(q(\bm{\theta}|\bm{\mu_\theta},\bm{\sigma_\theta})\|p(\bm{\theta}))\\
=&\sum_s \text{KL}(q(\theta_s|\mu_{\theta,s},\sigma_{\theta,s})\|p(\theta_s))\\
=&\frac{1}{2\overline{\sigma}_\theta^2}(\|\bm{\mu_\theta}\|^2+\|\bm{\sigma_\theta}\|^2)-\sum_s\log\left(\frac{\sigma_{\theta,i}}{\overline{\sigma}_\theta}\right) + \text{const}
\end{aligned}\end{equation*}
For the second part, we have
\begin{equation*}
\log p(D|\bm{\theta})=-\frac{1}{2\gamma_2^2}\sum_{i=1}^N\|\bm{f_\theta}({\y'}_i)-\y_i\|^2+\text{const}.
\end{equation*}
Combining above two equations together, proof is completed.
\subsection{Proof of Proposition \ref{prop:2}}\label{app:2}
This proof is based on literature \cite{vincent2011connection}. First, we have
 \begin{equation*}\begin{aligned}
&\mathbb{E}_{q(\widetilde{\x})}\left[\frac{1}{2}\left\|\bm{s}_{\bm{\phi}}(\widetilde{\x})-\frac{\partial \log q(\widetilde{\x})}{\partial \widetilde{\x}}\right\|^2\right]\\
=&\mathbb{E}_{q(\widetilde{\x})}\left[\frac{1}{2}\left\|\bm{s}_{\bm{\phi}}(\widetilde{\x})\right\|^2\right]-\mathbb{E}_{q(\widetilde{\x})}\left[\left\langle\bm{s}_{\bm{\phi}}(\widetilde{\x}),\frac{\partial \log q(\widetilde{\x})}{\partial \widetilde{\x}}\right\rangle\right]+c_1\\
\end{aligned}\end{equation*}
where $c_1$ denotes a constant. For the second term, we have
\begin{equation*}\begin{aligned}
&\mathbb{E}_{q(\widetilde{\x})}\left[\left\langle\bm{s}_{\bm{\phi}}(\widetilde{\x}),\frac{\partial \log q(\widetilde{\x})}{\partial \widetilde{\x}}\right\rangle\right]\\
=&\int q(\widetilde{\x})\left\langle\bm{s}_{\bm{\phi}}(\widetilde{\x}),\frac{\frac{\partial}{\partial \widetilde{\x}} q(\widetilde{\x})}{q(\widetilde{\x}) }\right\rangle \text{d}\widetilde{\x}\\
=&\int \left\langle\bm{s}_{\bm{\phi}}(\widetilde{\x}),\frac{\partial }{\partial \widetilde{\x}}\int q(\widetilde{\x}|\bm{\theta},\y)q(\bm{\theta})q(\y)\text{d}\bm{\theta}\text{d}\y\right\rangle \text{d}\widetilde{\x}\\
=&\int\left\langle\bm{s}_{\bm{\phi}}(\widetilde{\x}),\int q(\bm{\theta})q(\y)q(\widetilde{\x}|\bm{\theta},\y)\frac{\partial\log q(\widetilde{\x}|\bm{\theta},\y)}{\partial \widetilde{\x}}\text{d}\bm{\theta}\text{d}\y \right\rangle \text{d}\widetilde{\x}\\
=&\int q(\bm{\theta})q(\y)q(\widetilde{\x}|\bm{\theta},\y)\left\langle\bm{s}_{\bm{\phi}}(\widetilde{\x}),\frac{\partial\log q(\widetilde{\x}|\bm{\theta},\y)}{\partial \widetilde{\x}} \right\rangle \text{d}\widetilde{\x}\text{d}\bm{\theta}\text{d}\y\\
=&\mathbb{E}_{q(\widetilde{\x},\y,\bm{\theta})}\left[\left\langle\bm{s}_{\bm{\phi}}(\widetilde{\x}),\frac{\partial \log q(\widetilde{\x}|\bm{\theta},\y)}{\partial \widetilde{\x}}\right\rangle\right]
\end{aligned}\end{equation*}
where the first and fourth equations make use of the derivative property of the log function.
On the other hand, we have
 \begin{equation*}\begin{aligned}
&\mathbb{E}_{q(\widetilde{\x},\y,\bm{\theta})}\left[\frac{1}{2}\left\|\bm{s}_{\bm{\phi}}(\widetilde{\x})-\frac{\partial \log q(\widetilde{\x}|\bm{\theta},\y)}{\partial \widetilde{\x}}\right\|^2\right]\\
=&-\mathbb{E}_{q(\widetilde{\x},\y,\bm{\theta})}\left[\left\langle\bm{s}_{\bm{\phi}}(\widetilde{\x}),\frac{\partial \log q(\widetilde{\x}|\bm{\theta},\y)}{\partial \widetilde{\x}}\right\rangle\right]\\
&+\mathbb{E}_{q(\widetilde{\x})}\left[\frac{1}{2}\left\|\bm{s}_{\bm{\phi}}(\widetilde{\x})\right\|^2\right]+c_2\\
\end{aligned}\end{equation*}
where $c_2$ denotes a constant. The proof is completed.

\bibliographystyle{ieeetr}
\bibliography{library_manu}

\end{document}